\documentclass[a4paper,english,12pt]{article}

\usepackage[latin1]{inputenc}

\usepackage[english]{babel}

\usepackage{babel,amsmath,amssymb,amsbsy,amsfonts,latexsym,amsthm,mathrsfs}
\usepackage{graphicx}

\def \>{\rangle}

\def \<{\langle}

\begin{document}

\smallskip

\title{Beyond the constraints underlying Kolmogorov-Johnson-Mehl-Avrami theory related to the growth laws }
\author{M. Tomellini \\ Dipartimento di Scienze e Tecnologie Chimiche\\
 Universit\`{a} di Roma Tor Vergata\\
Via della Ricerca Scientifica 00133 Roma Italy  \\ \\ M. Fanfoni
\\ Dipartimento di Fisica Universit\`{a} di Roma Tor Vergata\\ Via
della Ricerca Scientifica 00133 Roma Italy}

\date{}

\maketitle

\begin{abstract}

The theory of Kolmogorov-Johnson-Mehl-Avrami (KJMA) for phase
transition kinetics is subjected to severe limitations concerning
the functional form of the growth law. This paper is devoted to
side step this drawback through the use of correlation function
approach. Moreover, we put forward an easy-to-handle formula,
written in terms of the experimentally accessible actual extended
volume fraction, which is found to match several types of growths.
Computer simulations have been done for corroborating the
theoretical approach.

\end{abstract}

\smallskip

\vskip 3cm

\  \

%%%%%%%%%%%%%%%%%%%%%%%%%%%%%%%%%%%%%%%%%%%%%%%%%%%

\newpage

\section{Introduction}

The Kolmogorov-Johnson-Mehl-Avrami (KJMA) model
\cite{{bib:K},{bib:J-M},{bib:A1-3}} finds application in a vast
ambit of scientific fields which ranges from Thin Film Growth to
Materials Science
\cite{{bib:Starink},{bib:Roura},{bib:Rikvold},{bib:Crespo},{bib:Bill},{bib:Korobov},{bib:Korobov2},{bib:Jcondmatrass}}
to Biology and Pharmacology \cite{{bib:DNA1Dim},{bib:Pharmacy}},
let alone the Applied Probability Theory
\cite{{bib:Moller},{bib:Erh}}. In the majority of these studies
the authors made use of a simplified version of the KJMA formula:
the stretched exponential $X(t)=1-\exp(-at^{n})$, where $X$ is the
fraction of the transformed phase, $a$ and $n$ (the latter known
as Avrami's exponent) being constants. The model, in principle, is
simple because it rests on a Poissonian stochastic process of
points in space, to which a growth law is attached. In fact, owing
to the Poissonian process, the nucleation takes place everywhere
in the space i.e. also in the already transformed phase. This
partially fictitious nucleation rate ($I(t)$), for we are dealing
with a Poissonian process, is linked to the actual (real)
nucleation rate ($I_{a}(t)$) according to:
$I_{a}(t)=I(t)[1-X(t)]$, where $X(t)$ is the transformed fraction.
The growth law transforms each point in a nucleus of radius
$R(t)$, $t$ stands for time. The pair, "points' generation" and
"growth law", is a key quantity of the theory. It happens that the
KJMA model fails for time-dependent points generation rate (i.e.
nucleation rate) associated with diffusional-type growth laws
\cite{bib:Shepilov2},\cite{bib:Aleks}. In particular, let us
define two classes of growth laws: i) $ d^{2}R/dt^{2}\geq0$ and
ii) $ d^{2}R/dt^{2}<0$. The KJMA model is suitable for describing
the first class of growths and for this reason we named it
KJMA-compliant as opposite to the second to which we attach the
adjective KJMA-non-compliant. The reason for that is due to the
particular stochastic process taken into account. As a matter of
fact, the Poissonian process requires that points can be generated
everywhere throughout the space independently of whether the space
is, because of growth, already transformed or not. Points
generated in the already transformed space are named phantoms
after Avrami. It goes without saying that, in the case of
KJMA-compliant growths, phantoms do not contribute to the true
transformed fraction, i.e. they are just virtual points whose only
role is to simplify the mathematics \cite{bib:Phantom1}. On the
other hand, in the KJMA-non-compliant growths phantoms may
contribute to the phase transition through the non-physical
"overgrowth" events \cite{bib:A1-3}. Incidentally, it is worth
noticing that the KJMA-non-compliant growths and KJMA-compliant
growths are indistinguishable if associated with simultaneous
nucleation \cite{bib:Ncemento}.

According to what has been said, one can summarize saying that the
concept of phantom implies the existence of the two classes of
growths.

One has to bring in mind that the above stretched exponential
expression ( $X(t)=1-\exp(-at^{n})$) is the exact solution of the
kinetics only in the case $I(t)=constant$ and $I(t)\sim\delta(t)$,
$I$ being the nucleation rate and $\delta(t)$ Dirac's delta
function, provided the growth is according to a power law. In
general, the term $at^{n}$ is a simple way to approximate the
convolution product between the nucleation rate (phantom included)
and the nucleus volume. This convolution is the "extended"
transformed fraction and takes into account the contribution of
phantoms, $\hat{X_{e}}$. In view of the large use of the KJMA
theory for dealing with experimental data, it should be desirable
to make use of an extended transformed fraction deprived of
phantom contribution, $X_{e}$. A pictorial view of the geometrical
meaning of $\hat{X_{e}}$, $X_{e}$ and $X$ is reported in fig.1.

\begin{figure}[htbp]
\begin{center}
\includegraphics[scale=2]{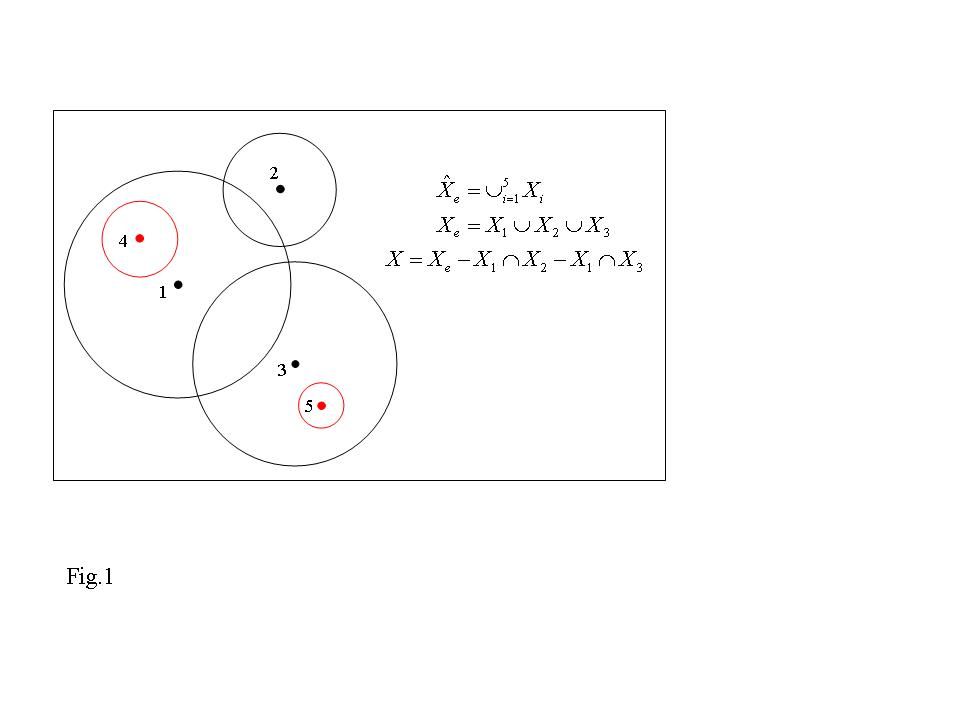}
\caption{Pictorial view of a random  ensemble
of  actual (dots 1-3) and phantom (dots 4-5) nuclei. The
definitions of both  phantom included ($\hat{X_{e}}$) and actual
($X_{e}$) extended fractions are also reported. The (extended)
volume of the i-th nucleus is denoted as $X_{i}$ and the
transformed fraction as $X$.}
\end{center}
\end{figure}

The aim of this contribution is twofold: i) to model the phase
transformation kinetics in terms of actual quantities such as the
nucleation rate; ii) to provide an expression for the transformed
fraction as a function of $X_{e}$.

\section{Theory}
In this section we discuss the stochastic theory that has to be
employed in order to get rid of phantoms in modeling the kinetics
of phase transformations ruled by nucleation and growth. To this
end, let us define the phantom included nucleation rate $I(t)$,
and the "actual" nucleation rate, $I_{a}(t)$, namely the rate of
birth of the "real" nuclei; $I_{a}(t)$ is the quantity
experimentally accessible. It goes without saying that a
mathematical formulation of the phase transition kinetics which
employs the "actual" nucleation rate holds true for both
KJMA-compliant and KJMA-non-compliant growth laws. This
automatically overcomes the limit of the KJMA approach. However,
actual nuclei imply a severe complication of the stochastic-nature
of the process under study, for we shift from a Poissonian to a
non-Poissonian process: in fact actual nuclei are spatially
correlated. Let us address this point in more detail by denoting
with $R(t,t')$ the radius of a nucleus, at running time $t$, which
starts growing at time $t'<t$. To be an actual nucleus it has to
lie at a distance $r>R(t',t'')$ from any other "older" nucleus
with $t''<t'$. In other words, in the spirit of the statistical
mechanics of hard spheres, this condition is formalized (at the
lowest order) through the pair distribution function for the "pair
of nuclei ($t',t''$)" at relative distance $r$

\begin{equation}\label{F1}
f_{2}(r,t',t'')\approx H(r-R(t',t'')),
\end{equation}

where $H(x)$ is the Heavyside function. Throughout the paper we
employ the notation by Van Kampen \cite{bib:VanKampen} according
to which $n-$dots distribution and correlation functions are
denoted as $f_{n}$ and $g_{n}$, respectively.

In previous papers, we have presented a theory for describing
phase transitions in the case of spatially correlated nuclei and
for time dependent nucleation rate
\cite{{bib:correlato2},{bib:correlato3}}. The untransformed
fraction can be expressed in terms of either the distribution
functions, ($f_{n}$-functions), or the correlation functions,
($g_{n}$-functions), where the $f_{n}$'s and $g_{n}$'s are linked
by cluster expansion \cite{bib:VanKampen}. Given a generic point
of space we have computed the probability that this point is not
covered (transformed) by any nucleus up to time $t$. This
probability is the fraction of untransformed phase i.e.
$Q(t)=1-X(t)$. By denoting with $\Delta(t,t')$ the volume of a
nucleus, which starts growing at time $t'<t$, at running time $t$,
and in the case of symmetric $f_{n}$ and $g_{n}$ functions, the
uncovered fraction is given by

\begin{eqnarray}\label{Uncov1}
Q(t) &=&
1-\int_{0}^{t}\tilde{I}(t_{1})dt_{1}\int_{\Delta(t,t_{1})}f_{1}(\mathbf{r}_{1})d\mathbf{r}_{1}+\\
\nonumber & + &
\int_{0}^{t}\tilde{I}(t_{1})dt_{1}\int_{0}^{t_{1}}\tilde{I}(t_{2})dt_{2}\int_{\Delta(t,t_{1})}d\mathbf{r}_{1}\int_{\Delta(t,t_{2})}d\mathbf{r}_{2}f_{2}(\mathbf{r}_{1},\mathbf{r}_{2})-....\\
\nonumber & = & 1+\\ \nonumber & + &
\sum_{m=1}^{\infty}\frac{(-1)^{m}}{m!}\int_{0}^{t}\tilde{I}(t_{1})dt_{1}..\int_{0}^{t}\tilde{I}(t_{m})dt_{m}\times
\nonumber  \\ \nonumber & & \:
\int_{\Delta(t,t_{1})}d\mathbf{r}_{1}\int_{\Delta(t,t_{2})}d\mathbf{r}_{2}....\int_{\Delta(t,t_{m})}f_{m}(\mathbf{r}_{1},..,\mathbf{r}_{m})d\mathbf{r}_{m}
\end{eqnarray}

or,
\begin{eqnarray}\label{Uncov2}
Q(t) &=& \exp\Big[\sum_{m=1}^{\infty}\frac{(-1)^{m}}{m!}\int_{0}^{t}\tilde{I}(t_{1})dt_{1}...\int_{0}^{t}\tilde{I}(t_{m})dt_{m} \nonumber  \\
& & \:
\int_{\Delta(t,t_{1})}d\mathbf{r}_{1}\int_{\Delta(t,t_{2})}d\mathbf{r}_{2}....\int_{\Delta(t,t_{m})}g_{m}(\mathbf{r}_{1},..,\mathbf{r}_{m})d\mathbf{r}_{m}\Big].
\end{eqnarray}

It is worth pointing out that the nucleation rates entering these
equations are in fact subjected to the condition imposed by the
correlation among nuclei. This quantity may or may not imply
phantoms depending on the specific form of the $f_{n}$ functions.
For this reason we introduce the new symbol $\tilde{I}$. In
particular, for the hard core correlation (eqn.\ref{F1})
$\tilde{I}$ coincides with the actual nucleation rate,
$\tilde{I}=I_{a}$, which leads to the solution of the phase
transition kinetics in terms of the actual nucleation rate.

Since eqn.\ref{Uncov1},\ref{Uncov2} are the exact solutions of the
stochastic process linked to the phase transition, they also
coincide with the KJMA formula provided the above mentioned
preconditions are met i.e. random nucleation and KJMA-compliant
growth.

As far as the hard-core correlation (eqn.\ref{F1}) and the
kinetics eqns.\ref{Uncov1},\ref{Uncov2} are concerned, we note
that the number of nuclei of size $R(t,t_{1})$ is
$I_{a}(t_{1})dt_{1} =O(dt_{1})$. As a consequence and in the
framework of the statistical mechanics of hard sphere fluid, we
are dealing with an extremely dilute solution of pairs of
components $"t_{1}, t_{2}"$. Accordingly, the number density of
$t_{2}$ spheres being of the order of $O(dt_{2})$, higher order
terms in the cluster expansion of the $f_{2}$ function can be
neglected, thus $f_{2}(r,t',t'')=H(r-R(t',t''))$ in
eqn.\ref{Uncov1}.

\subsection{KJMA-compliant growths}

In this section we show that eqn.\ref{Uncov1} is compatible with
the KJMA kinetics only in the case of KJMA-compliant function. On
the other hand, such a comparison will give a deeper insight into
the reasons why the KJMA kinetics does not work in the case of
"KJMA-non-compliant" functions. In the following, we discuss the
linear growth law for 2D case. Also, to simplify the complexity of
the computation the "actual" nucleation rate, $I_{a}$, is taken as
constant; as a consequence the phantom included nucleation rate
reads \cite{{bib:A1-3},{bib:Ncemento}}
$I(t)=I_{a}/(1-X(t))=I_{a}/Q(t)\equiv I_{a} F(t)$ and the KJMA
kinetics becomes

\begin{equation}\label{KJMA2}
F(t)=\exp\Big[\int_{0}^{t}I_{a}F(t')\pi R^{2}(t,t')dt'\Big],
\end{equation}

where $R(t,t')=v(t-t')$ and $v$  is a constant.

We consider the series expansion of $F(t)$ around $t=0$. One gets,

\begin{equation}\label{serie1}
\frac{d^{n+1}F(t)}{dt^{n+1}}\equiv
F^{(n+1)}=(F\Omega)^{(n)}=\sum_{k=0}^{n}(^{n}_{k})F^{(n)}\Omega^{(n-k)},
\end{equation}

where $\Omega(t)=2\pi
I_{a}\int_{0}^{t}F(t')R(t,t')\partial_{t}R(t,t')dt'=2\pi
I_{a}v\int_{0}^{t}F(t')R(t,t')dt'$.

Moreover, since $\Omega^{(m+2)}=\kappa F^{(m)}$, being
$\kappa=2\pi v^{2}I_{a}$ and $F^{(0)}=1,
\Omega^{(0)}=\Omega^{(1)}=0$, from eqn.\ref{serie1} it is found
that only the terms $F^{(3n)}(0)$ are different from zero, i.e.
$F(t)=\sum_{n=0}^{\infty}\frac{1}{(3n)!}F^{(3n)}(0)t^{3n}=\sum_{m=0}^{\infty}c_{m}t^{m}$.
In particular, the first coefficients are: $F^{(3)}(0)=\kappa,
F^{(6)}(0)=11\kappa^{2}, F^{(9)}(0)=375\kappa^{3},
F^{(12)}(0)=234147\kappa^{4}$.

Next we derive the series expansion of the untransformed fraction
$Q(t)=\sum_{n=0}^{\infty}b_{n}t^{n}$, by exploiting the condition
$1=Q(t)F(t)=\sum_{n=0}^{\infty}b_{n}t^{n}\sum_{m=0}^{\infty}c_{m}t^{m}$.
Even in this case the $b_{n}$ coefficients are different from zero
for $n=3k$ (with integer $k$). The first four coefficients are

\begin{equation}\begin{array}{l}\label{coeffic1}
b_{0}=\frac{1}{c_{0}}=1\\
b_{3}=-c_{3}\\ b_{6}=-c_{6}+c_{3}^{2}\\
b_{9}=-c_{9}+2c_{3}c_{6}-c_{3}^{2}.
\end{array}
\end{equation}

By using the expression of $c_{n}$'s the series of the
untransformed fraction up to $t^{9}$ is given as

\begin{equation}\begin{array}{l}\label{serie2}
Q(t)=1-\frac{1}{6}\kappa t^{3}+ \frac{1}{80}\kappa^{2}
t^{6}-\frac{207}{9!}\kappa^{3} t^{9}+O(t^{12}).
\end{array}
\end{equation}

Since $X_{e}=\frac{1}{3!}\kappa t^{3}=\frac{1}{3}\pi
I_{a}v^{2}t^{3}=\int_{0}^{t}I_{a}\pi R^{2}(t,t')dt'$ the series
eqn.\ref{serie2} can be rewritten as

\begin{equation}\label{serie3}
Q(t)=1-X_{e}+\frac{9}{20}X_{e}^{2}-\frac{69}{560}X_{e}^{3}+
O(X_{e}^{4}),
\end{equation}

the two last coefficients being $0.45$ and $0.123$. We emphasize
that the coefficients of this series only depend upon growth law,
for constant $I_{a}$ (see also the last section).

The next step is to show that the untransformed fraction, given by
the series eqn.\ref{Uncov1} is equal to eqn.\ref{serie2} or
eqn.\ref{serie3}. We have carried out the first two terms,
exactly, while, owing to the tremendous computational complexity,
for the third term an approximation has been employed. It is worth
reminding that the distribution functions are
$f_{1}(\mathbf{r}_{1})=1,
f_{2}(\mathbf{r}_{12},t_{1},t_{2})=H(r_{12}-R(t_{1},t_{2}))$ and
$f_{3}(\mathbf{r}_{12},\mathbf{r}_{13},\mathbf{r}_{23},t_{1},t_{2},t_{3})=
H(r_{12}-R(t_{1},t_{2}))H(r_{13}-R(t_{1},t_{3}))H(r_{23}-R(t_{2},t_{3}))$,
where $\mathbf{r}_{ij}$ is the relative distance. In fact, since
$I_{a}dt_{i}=O(dt_{i})$ the system of dots is dilute and
$f_{3}(1,2,3)=f_{2}(1,2)f_{2}(1,3)f_{2}(3,2)$ i.e. the
superposition principle holds true \cite{bib:Hill}. Since the
system is homogeneous the $f_{n}$ functions depend on
$r_{ij}=|\mathbf{r}_{ij}|=|\mathbf{r}_{i}-\mathbf{r}_{j}|$.
Eqn.\ref{Uncov1} becomes

\begin{equation}\begin{array}{c}\label{coeffic1}
Q(t)=1-I_{a}\int_{0}^{t}dt_{1}\int_{\Delta(t,t_{1})}d\mathbf{r}_{1}+\\ \\+I_{a}^{2}\int_{0}^{t}dt_{1}\int_{0}^{t_{1}}dt_{2}\int_{\Delta(t,t_{1})}d\mathbf{r}_{1}\int_{\Delta(t,t_{2})}d\mathbf{r}_{2}H(r_{12}-R(t_{1},t_{2}))+\\ \\
-I_{a}^{3}\int_{0}^{t}dt_{1}\int_{0}^{t_{1}}dt_{2}\int_{0}^{t_{2}}dt_{3}\int_{\Delta(t,t_{1})}d\mathbf{r}_{1}\int_{\Delta(t,t_{2})}d\mathbf{r}_{2}\times\\ \\
\int_{\Delta(t,t_{3})}d\mathbf{r}_{3}H(r_{12}-R(t_{1},t_{2}))H(r_{23}-R(t_{2},t_{3}))H(r_{13}-R(t_{1},t_{3}))+...,
\end{array}
\end{equation}

where the integration domain $\Delta(t,t_{i})$ is the circle of
radius $R(t,t_{i})=v(t-t_{i})$. The $f_{1}$ containing term  is
the extended surface fraction
$X_{e}(t)=I_{a}\int_{0}^{t}dt_{1}|\Delta(t,t_{1})|$ and coincides
with the second term of the expansion eqn.\ref{serie3}. Let us
focus our attention on the integrals in the spatial domain - for
the sake of clarity shown in fig.2a together with the circle of
correlation $R(t_{1},t_{2})$- of the $f_{2}$ containing term.  By
employing relative coordinates the integrals read

\begin{figure}[htbp]
\begin{center}
\includegraphics[scale=2]{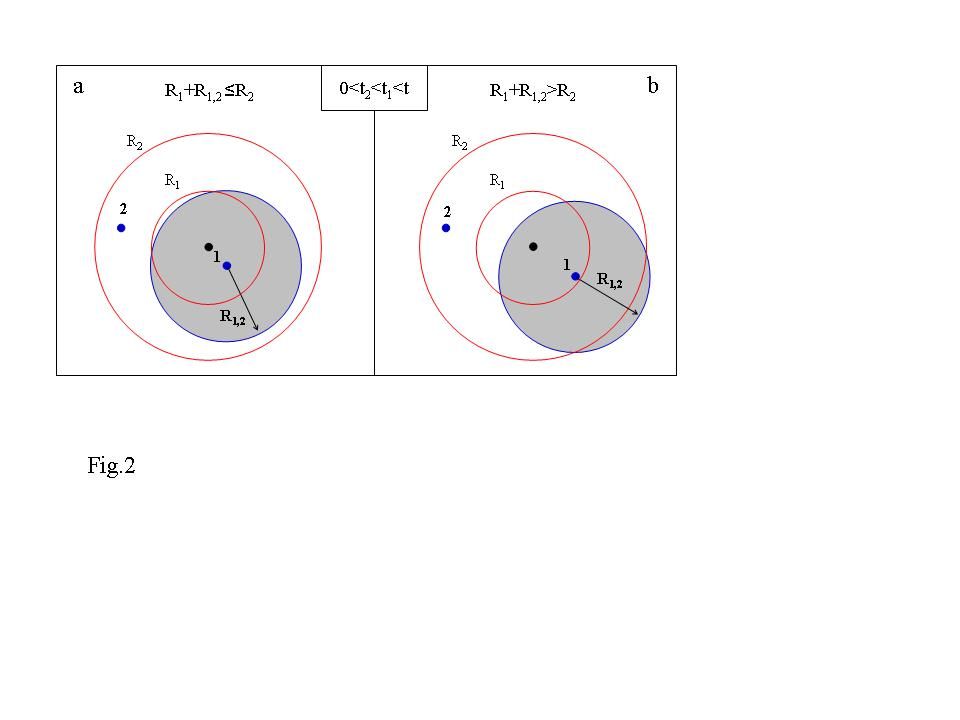}
\caption{The integration domains and the
correlation circles are depicted for the integrals over the
$f_{2}$ function. The cases of KJMA-compliant and
KJMA-non-compliant growths are reported in panels a and b,
respectively. In the drawing $R_{12}\equiv R(t_{1},t_{2})$,
$R_{i}\equiv R(t,t_{i})$ ($i=1,2$) and $t_{2}<t_{1}$. In the case
of KJMA-compliant growth $R_{1}+R_{12}\leq R_{2}$ and the hard
core circle is within the integration domain, $R_{2}$ (a). In the
case of KJMA-non-compliant growths $R_{1}+R_{12}>R_{2}$ and the
hard core disk overcomes the integration domain of the "older
nucleus" (b).
}
\end{center}
\end{figure}

\begin{equation}\label{Area0}
\int_{\Delta(t,t_{1})}d\mathbf{r}_{1}
\int_{\Delta(t,t_{2})}d\mathbf{r}_{12}H(r_{12}-R(t_{1},t_{2}))=\int_{\Delta(t,t_{1})}d\mathbf{r}_{1}A(r_{1},t,t_{1},t_{2}),
\end{equation}

where $A(r_{1},t,t_{1},t_{2})=A(r_{1},R(t,t_{2}),R(t_{1},t_{2}))$
is the area "spanned" by the second nucleus when the first one is
located at $\mathbf{r}_{1}$ ($t_{2}<t_{1}<t$). It is at this point
of the computation that the growth law comes into play; indeed in
the case of KJMA-compliant growth laws (here linear growth) the
correlation circle $R(t_{1},t_{2})$ is entirely within the
integration domain $R(t,t_{2})$ (fig.2a). Consequently,

\begin{equation}\label{Aria1}
A(t,t_{1},t_{2})=\pi \big[ R^{2}(t,t_{2})-R^{2}(t_{1},t_{2}) \big]
\end{equation}

is independent of $\mathbf{r}_{1}$. On the other hand, in the case
of KJMA-non-compliant growth laws the correlation circle overcomes
the integration domain of the second nucleus and the relationship
above does not hold true, anymore (see fig.2b). In general, the
KJMA-compliant functions satisfy the condition
$R(t,t_{1})+R(t_{1},t_{2})\leq R(t,t_{2})$ i.e. for a power growth
law $(t-t_{1})^{n}+(t_{1}-t_{2})^{n}\leq (t-t_{2})^{n}$, which is
verified only for $n\geq 1$. In fact, by setting
$\tau=t_{1}-t_{2}$ and $\eta = \frac{t-t_{2}}{\tau}>1$, the
inequality above reads $( \eta -1)^{n}\leq \eta^n -1$, which is
satisfied for $n\geq 1$. On the other hand, for $n=1/k$ (with
integer $k>1$) the inequality is  $[(\eta -1)^{1/k} +1 ]^{k}\leq
\eta$ namely,

\begin{equation}\label{Growth1}
\sum _{\mu =1}^{k-1}(^{k}_{\mu})(\eta -1)^{\frac{\mu}{k}}\leq0,
\end{equation}

which is never satisfied ($\eta>1$).

For the KJMA-compliant growth the  contribution of the $f_{2}$
containing term becomes

\begin{equation}\begin{array}{c}\label{Secondterm}
\pi
I_{a}^{2}\int_{0}^{t}dt_{1}\int_{0}^{t_{1}}dt_{2}\int_{\Delta(t,t_{1})}d\mathbf{r}_{1}
\big[ R^{2}(t,t_{2})-R^{2}(t_{1},t_{2})
\big]\\ \\
=\pi^{2}I_{a}^{2}\int_{0}^{t}dt_{1}\int_{0}^{t_{1}}dt_{2}R^{2}(t,t_{1})
\big[ R^{2}(t,t_{2})-R^{2}(t_{1},t_{2})\big]\\ \\
=\frac{X_{e}^{2}}{2}-\pi^{2}I_{a}^{2}\int_{0}^{t}dt_{1}\int_{0}^{t_{1}}dt_{2}R^{2}(t,t_{1})R^{2}(t_{1},t_{2}).
\end{array}
\end{equation}

It is possible to show that for linear growth the last term of
eqn.\ref{Secondterm} is equal to $\frac{1}{180}(\pi
I_{a}v^{2}t^{3})^{2}$, consequently we get
$\frac{X_{e}^{2}}{2}-\frac{9X_{e}^{2}}{180}=\frac{9}{20}X_{e}^{2}$,
which coincides with the term of the same order in the KJMA series
eqn.\ref{serie3}.

Let's now briefly consider the contribution of the $f_{3}$
containing term in the general expression eqn.\ref{Uncov1}. Fig.3
shows the integration domains $R_{i}=R(t,t_{i})$ and the
correlation circles $R(t_{i},t_{j})$ for the three nuclei born at
time $t_{i}$ ($i=1,2,3$).  The configuration integral over $f_{3}$
in eqn.\ref{coeffic1} becomes

\begin{figure}[htbp]
\begin{center}
\includegraphics[scale=2.2]{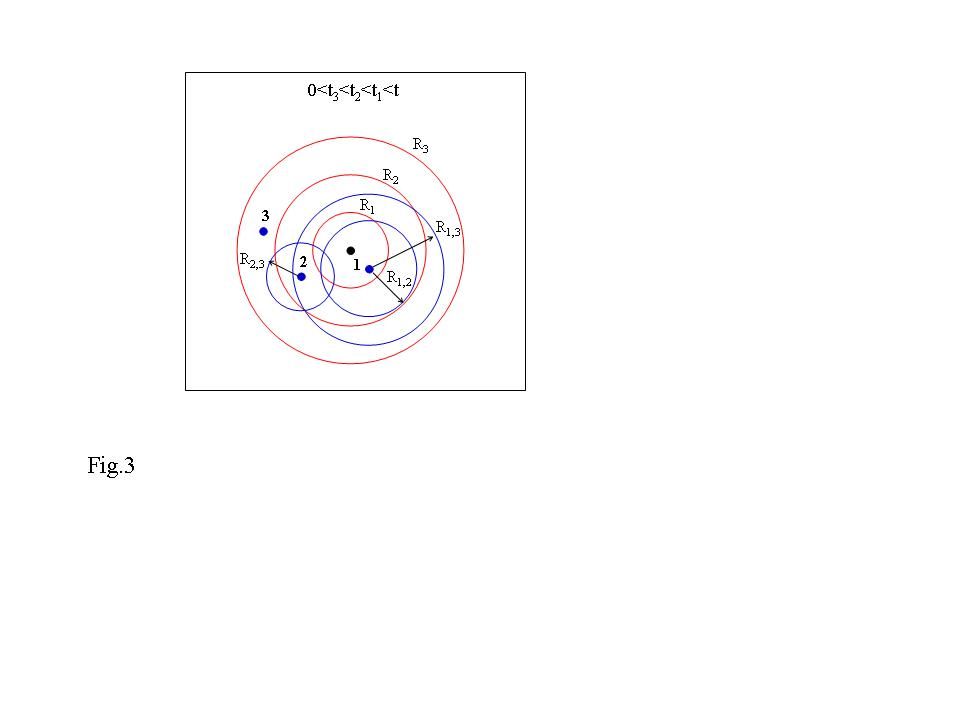}
\caption{Integration domains and correlation
hard disks for the integral over the $f_{3}$ function. In the
drawing $R_{ij}\equiv R(t_{i},t_{j})$ and $R_{i}\equiv
R(t,t_{i})$. In the case of KJMA-compliant growth all the circles
are entirely within the $R(t,t_{3})$ circle.}
\end{center}
\end{figure}

\begin{equation}\begin{array}{l}\label{f3}
I_{a}^{3}\int_{0}^{t}dt_{1}\int_{0}^{t_{1}}dt_{2}\int_{0}^{t_{2}}dt_{3}\int_{\Delta(t,t_{1})}d\mathbf{r}_{1}\int_{\Delta(t,t_{2})}d\mathbf{r}_{2}H(r_{12}-R(t_{1},t_{2}))\times\\
\\\int_{\Delta(t,t_{3})}d\mathbf{r}_{3}H(r_{13}-R(t_{1},t_{3}))H(r_{23}-R(t_{2},t_{3}))\\
\\
=I_{a}^{3}\int_{0}^{t}dt_{1}\int_{0}^{t_{1}}dt_{2}\int_{0}^{t_{2}}dt_{3}
\times\\
\\\int_{\Delta(t,t_{1})}d\mathbf{r}_{1}\int_{\Delta(t,t_{2})}d\mathbf{r}_{2}H(r_{12}-R(t_{1},t_{2}))A(r_{12},R(t_{1},t_{3}),R(t_{2},t_{3})),
\end{array}
\end{equation}

where $t_{3}<t_{2}<t_{1}<t$ is assumed. In this equation,
$A(r_{12},R(t_{1},t_{3}),R(t_{2},t_{3}))$ is the area spanned by
the third nucleus when the first and the second are located at
$\mathbf{r}_{1}$ and $\mathbf{r}_{2}$, respectively. Because of
the possible overlap between the correlation circles
$R(t_{1},t_{3})$ and $R(t_{2},t_{3})$, and for $R(t_{1},t_{3})$ is
encompassed within $R(t,t_{3})$ this area is a function of
$r_{12}$ as

\begin{equation}\begin{array}{l}\label{Areapalinearf3}
A(r_{12})=\omega(r_{12})
H(R(t_{1},t_{3})+R(t_{2},t_{3})-r_{12})+\\ \\ +
 \pi\big[ R^{2}(t,t_{3})-R^{2}(t_{1},t_{3})-R^{2}(t_{2},t_{3})
\big]H(r_{12}-R(t_{2},t_{3})-R(t_{1},t_{3})),
\end{array}
\end{equation}

where $\omega(r_{12})=\pi
(R^{2}(t,t_{3})-R^{2}(t_{1},t_{3})-R^{2}(t_{2},t_{3}))+\varpi(r_{12},R(t_{1},t_{3}),R(t_{2},t_{3}))$,
with $\varpi(x,\rho_{1},\rho_{2})$ being the overlap area of two
circles of radius $\rho_{1}$ and $\rho_{2}$ at relative distance
$x$

\begin{equation}\begin{array}{l}\label{Areoverlap}
\varpi(x,\rho_{1},\rho_{2})=
-\frac{1}{2}\sqrt{4x^{2}\rho_{1}^{2}-[\rho_{2}^{2}-x^{2}-\rho_{1}^{2}
 ]^{2}}+\\ \\ +\rho_{1}^{2}\arccos
\frac{\rho_{1}^{2}+x^{2}-\rho_{2}^{2}}{2x\rho_{1}}+
\rho_{2}^{2}\arccos\frac{\rho_{2}^{2}+x^{2}-\rho_{1}^{2}}{2x\rho_{2}}.
\end{array}
\end{equation}

It turns out that the computation of the third order term of the
series, eqn.\ref{f3}, is a formidable task indeed. We do not
attempt to perform the exact estimate of this term which, however,
must coincide with the same order term of the KJMA series. On the
other hand, an approximate evaluation of this term by using an
oversimplified form of the $A(r_{12})$ area, is possible by
formally rewriting this area as $ A(r_{12})=\pi
[R^{2}(t,t_{3})-R^{2}(t_{1},t_{3})-\beta R^{2}(t_{2},t_{3})] $,
where $\beta \in (0,1)$ is given by
$\beta=[1-\frac{\varpi(r_{12})}{ \pi R^{2}(t_{2},t_{3})}
H(R(t_{1},t_{3})+R(t_{2},t_{3})-r_{12})]$. In the case of
complete-overlap ($\beta=0$) we get,

\begin{equation}\begin{array}{l}\label{f3approx}
\pi^{3}I_{a}^{3}\int_{0}^{t}dt_{1}\int_{0}^{t_{1}}dt_{2}\int_{0}^{t_{2}}dt_{3}R^{2}(t,t_{1})[R^{2}(t,t_{2})-R^{2}(t_{1},t_{2})][R^{2}(t,t_{3})-R^{2}(t_{1},t_{3})],

\end{array}
\end{equation}
that for the linear growth ($R(t,t')=v(t-t')$) gives
$\frac{24}{7!}I_{a}^{3}\pi^{3}v^{6}t^{9}=\frac{72}{560}X_{e}^{3}$
to be compared with the exact value $\frac{69}{560}X_{e}^{3}$
which brings an uncertainty of $4.3\%$ .

\subsection{KJMA-non-compliant growths}

Let us now consider the parabolic growth $R(t,t')=v\sqrt{t-t'}$.
In this case the series expansion of the function  $F(t)=1/Q(t)$,
given by eqn.\ref{KJMA2}, can be performed by employing the same
computation pathway discussed above, where now
$\Omega^{(n+1)}=\kappa F^{(n)}$ and $\kappa=\pi I_{a}v^{2}$. In
this case $F^{(2n)}(0)\neq 0$ implying
$F(t)=1+\frac{\kappa}{2}t^{2}+\frac{1}{6}\kappa^{2}t^{4}+\frac{34}{6!}\kappa^{3}t^{6}$
and

\begin{equation}\begin{array}{c}\label{parabola1}
Q(t)=1-\frac{\kappa}{2}t^{2}+\frac{\kappa^{2}}{12}t^{4}-\frac{\kappa^{3}}{180}t^{6}+O(t^{8})\\
\\=1-X_{e}+\frac{1}{3}X_{e}^{2}-\frac{4}{90}X_{e}^{3}+O(X_{e}^{4}),
\end{array}
\end{equation}

the two last coefficients being $0.33$ and $0.044$. It is worth
pointing out that in such an evaluation the transformed fraction,
$X$, is comprehensive of the contribution of phantoms. In fact, we
recall that eqn.\ref{KJMA2} is the KJMA solution with the phantom
included nucleation rate, $I_{a}/(1-X)$.

 The $f_{1}$ containing term of eqn.\ref{Uncov1} gives the extended
volume fraction $\frac{\kappa}{2}t^{2}$. As far as the third term
is concerned ($f_{2}$ contribution), it is possible to show that
also for $n=1/2$ the integral eqn.\ref{Secondterm} coincides with
the third term of eqn.\ref{parabola1},
$\frac{\kappa^{2}}{12}t^{4}$. However, it is important to stress
that eqn.\ref{Secondterm} does not coincide, in this case, with
the integral over the $f_{2}$ function of the exact solution
eqn.\ref{coeffic1}, since in the latter equation enter the actual
nuclei, only. From the mathematical point of view, in the case of
parabolic growth, the correlation circle is not contained within
the integration domain as depicted in Fig.2b. In other words, for
KJMA-non-compliant growth laws the area $A$ is a function of
$r_{1}$ and the term of order $I_{a}^{2}$ in eqn.\ref{coeffic1}
does not coincide with $\frac{X_{e}^{2}}{3}$ of
eqn.\ref{parabola1} (parabolic growth). In particular, under these
circumstances we get

\begin{equation}\begin{array}{c}\label{Areaparabolic}
A(r_{1},t,t_{1},t_{2})=\omega(r_{1})
H(r_{1}+R(t_{1},t_{2})-R(t,t_{2}))+\\ \\+
 \pi\big[ R^{2}(t,t_{2})-R^{2}(t_{1},t_{2})
\big]H(R(t,t_{2})-r_{1}-R(t_{1},t_{2})),
\end{array}
\end{equation}

where $\omega(r_1)=\pi R^{2}(t,t_{2})-\varpi(r_{1},
R(t,t_{2}),R(t_{1},t_{2}))$ with $\varpi(r_{1}) $ the overlap area
of two circles of radius $R(t,t_{2})$ and $R(t_{1},t_{2})$ at
distance $r_{1}$ (eqn.\ref{Areoverlap}).

\section{Numerical Simulations}

The ultimate aim of this section is to propose a simple formula
for describing the kinetics on the basis of the "actual" extended
transformed fraction, $X_{e}$. On the ground of eqn.\ref{Uncov2}
the transformed fraction can be rewritten in the general form

\begin{equation}\label{sim1}
X(t)=1-\exp[-X_{e}(t)\gamma (X_{e}(t))],
\end{equation}

where $\gamma(X_{e}(t))$ embodies the contributions of
correlations among nuclei \cite{bib:correlato2}. It is worth
noticing that for KJMA-complaint growths (random nucleation)
eqn.\ref{Uncov2} actually  reproduces the KJMA formula. In fact by
identifying $\tilde{I}$ with the phantom included nucleation rate
( $\tilde{I}\equiv I$) one gets $g_{m>1}=0$ leading to the formula
$Q=\exp(-\hat{X_{e}})$. On the other hand, working with the actual
nucleation rate, in eqn.\ref{Uncov2} $\tilde{I}\equiv I_{a}$,
$g_{m}\neq0$ and the series has infinite terms. $\gamma
(X_{e}(t))$ can therefore be expanded as a power series of the
extended actual volume fraction, $X_{e}$. Moreover, by exploiting
the homogeneity properties of the $f_{n}$ functions (see below),
it is possible to attach a physical meaning to the power series
coefficients, in terms of nucleation rate and growth law. Also,
for constant $I_{a}$ the coefficients of this series only depend
upon growth law. To the aim of achieving a suitable compromise
between handiness and pliability, we retain the liner
approximation of $\gamma (X_{e}(t))$ obtaining the following
kinetics

\begin{equation}\label{sim2}
X=1-\exp[-(aX_{e}+bX_{e}^{2})],
\end{equation}

with $a$ and $b$ constants. For the sake of completeness we point
out that, according to its physical meaning, the parameter $a$
should be unitary. Nevertheless, the substitution of the infinite
expansion with only two terms authorizes the introduction of the
new parameter $a$. In any case, the $a$ values are found to be
nearly one (see fig.7 below).

In order to study the transition kinetics in terms of actual
nuclei and to test eqn.\ref{sim2}, we worked out 2D computer
simulations for several growth laws at constant nucleation rate,
$I_{a}$.

As typical for this kind of study \cite{bib:Roura},
\cite{bib:Sessa}, the simulation is performed on a lattice (square
in our case) where, in order to mimic the continuum case, the
lattice space is much lower than the mean size of nuclei.

In particular, the transformation takes place on a square lattice
whose dimension is $1000\times 1000$ with a nucleation rate of
$I_{a}=3$. It is worth reminding that, since the nucleation is
Poissonian, it occurs on the entire lattice independently of
whether the space is already transformed or not. The computer
simulation can be run taking into account the presence of phantoms
or not. In the former case the outputs have been labeled as "w"
while the latter as "wo". As far as the growth laws are concerned,
we limited ourself to the power laws, $R(t)\sim t^{n}$ for $n=1/4,
1/3, 1/2, 1, 3/2$ and $2$.

The results of the simulations are displayed in figs.4a-c for the
KJMA-non-compliant growths. In particular, the fractional surface
coverage,$X$, as a function of the actual extended fraction
$X_{e}$, with and without the contribution of phantoms, are
reported (curves labeled with "w" and "wo", respectively).

\begin{figure}[htbp]
\begin{center}
\includegraphics[scale=1.5]{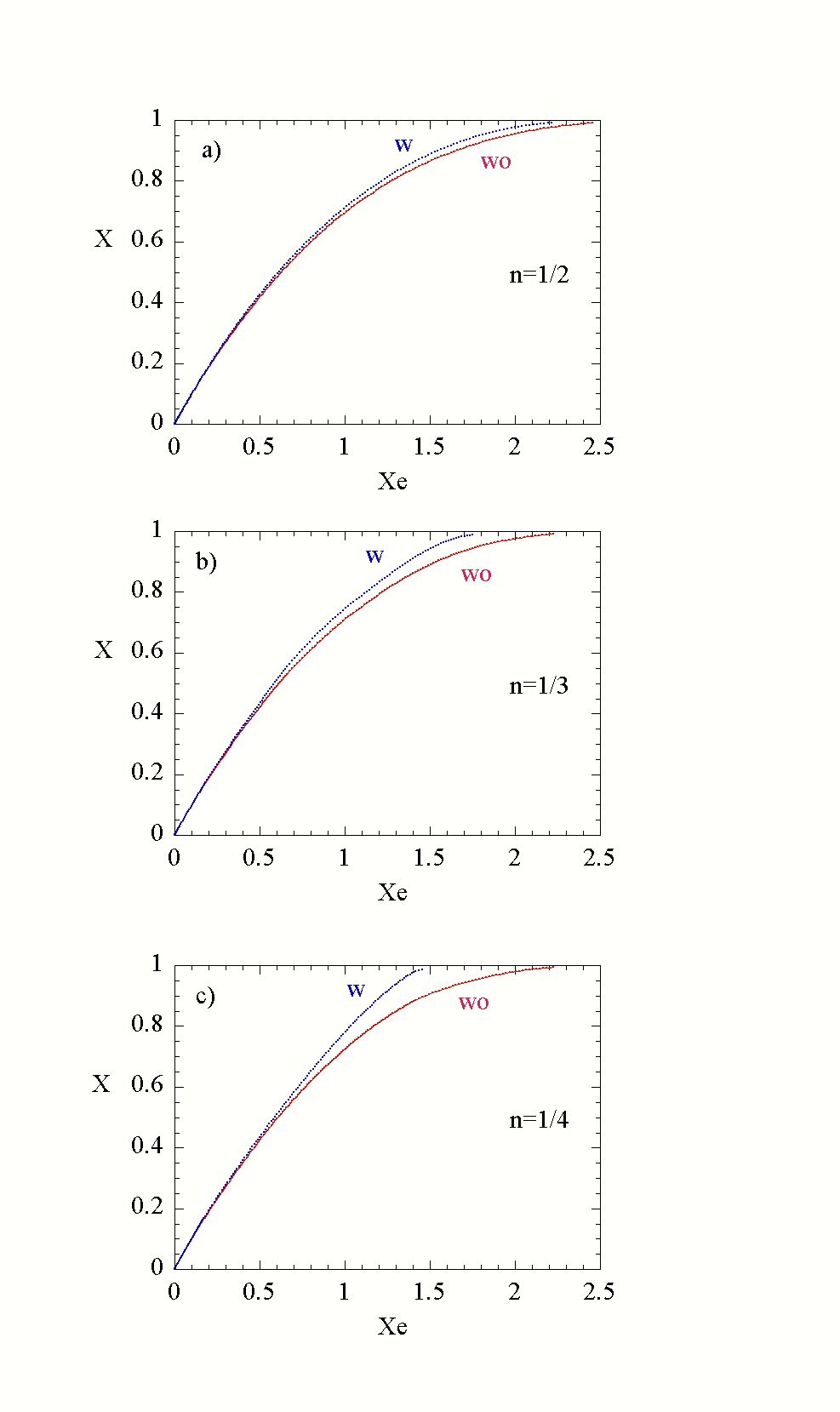}
\caption{Computer simulatiosn of phase
transformations ruled by KJMA-non-compliant growths. The surface
fraction, $X$, is shown as a function of the extended fraction
$X_{e}$, for the power law ($R\approx t^{n}$) where $n=1/2$, $1/3$
and $1/4$ in figures a, b and c, respectively. The kinetics  with
(w) and without (wo) the inclusion of phantoms  are displayed.}
\end{center}
\end{figure}

 The contribution of phantom overgrowth to the transformation kinetics
is highlighted in fig.5 and shows that this effect brings an
uncertainty on $X$ which ranges from  $2\%$ to $5\%$ on going from
$n=1/2$ to $n=1/4$. In the case of parabolic growth this figure is
lower than $2\%$. 

\begin{figure}[htbp]
\begin{center}
\includegraphics[scale=1]{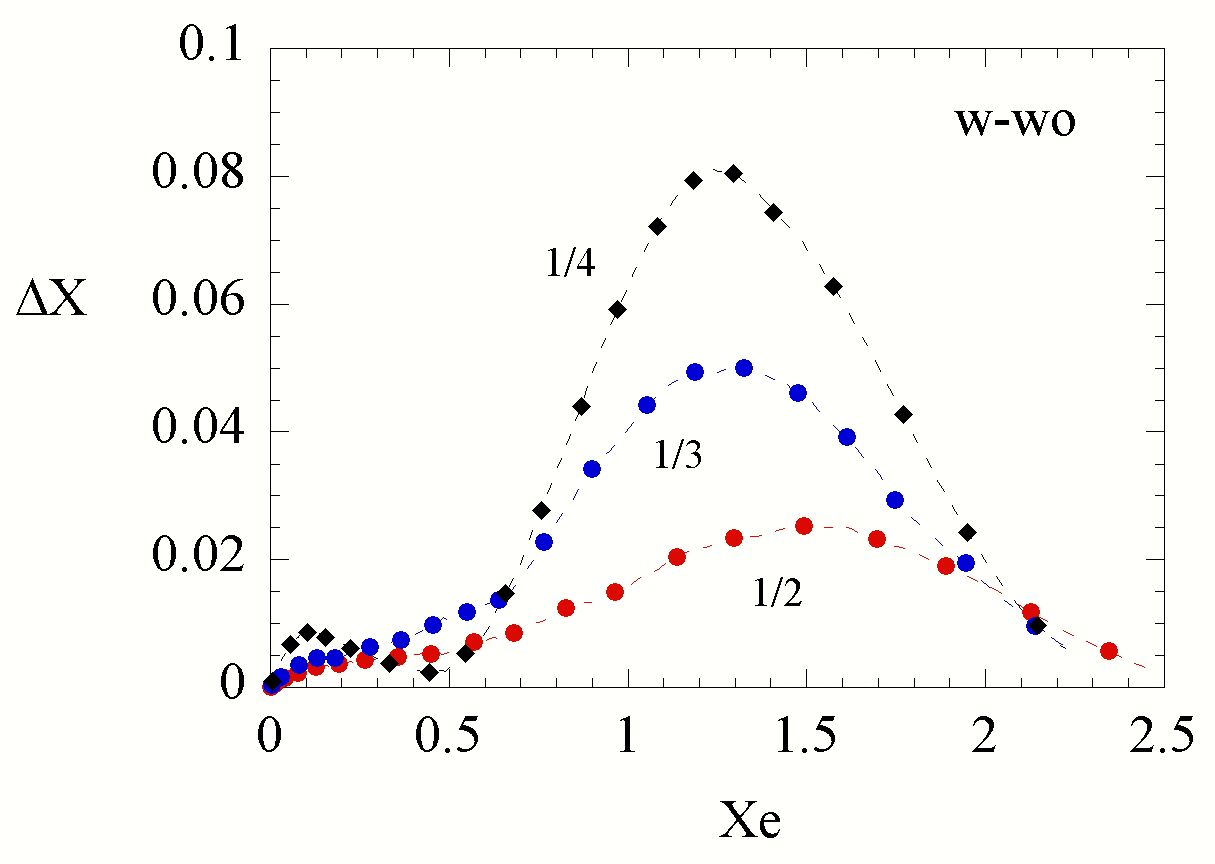}
\caption{Contribution of phantom overgrowth
to the kinetics reported in fig.4 (displayed is the difference
between curves (w) and (wo)). The area beneath the curves
normalized to the area of the kinetics of fig.4 are, respectively,
$0.053$, $0.038$ and $0.02$ for $n=1/4$, $n=1/3$ and $n=1/2$.}
\end{center}
\end{figure}

These results are in qualitative agreement with
previous studies on phantom overgrowth, although performed for a
different nucleation laws \cite{{bib:Sessa},{bib:Shepilov2}}. As
discussed in more details below, the results displayed in figs.4,5
are universal, i.e. they only depend on power exponent, $n$, and
nucleation law (in the present case $I_{a}=constant$).
Accordingly, the lower $n$ the more important is phantom
overgrowth. In fact, let us  consider a phantom, which starts
growing at time $\bar{t}$, located at $r_{p}$ from the center of
an actual nucleus which starts growing at $t=0$ (fig.6a). For
KJMA-non-compliant growth, $R(t-t')=v(t-t')^{\frac{1}{k}}$ (with
integer $k>1$), the phantom overtakes the actual nucleus at time
$t_{o}$, that is the solution of the equation
$r_{p}+v(t_{o}-\bar{t})^{\frac{1}{k}}=vt_{o}^{\frac{1}{k}}$,
namely

\begin{figure}[htbp]
\begin{center}
\includegraphics[scale=1.8]{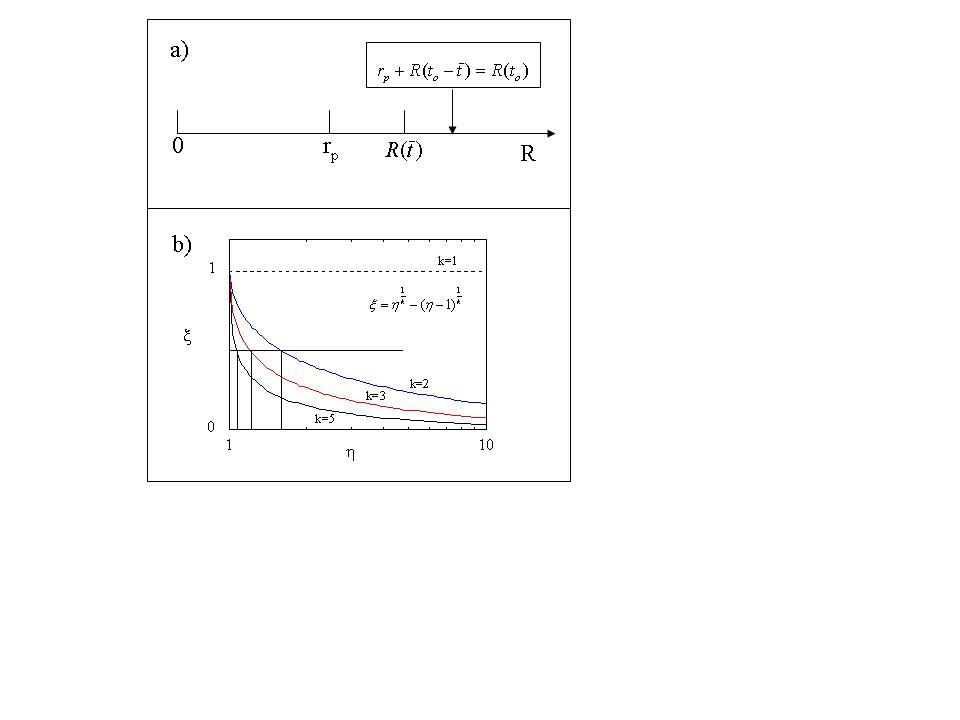}
\caption{Sketch of the overgrowth process
in the space domain. $r_{p}$ denotes the location of the phantom
which start growing at time $\bar{t}$ when the size of the actual
nucleus is $R(\bar{t})$. The phantom overtakes the actual nucleus
at time $t_{o}$ when the size of the actual nucleus is $R(t_{o})$.}
\end{center}
\end{figure}

\begin{equation}\label{ovegrowth}
\xi = \eta ^{\frac{1}{k}}-(\eta-1)^{\frac{1}{k}},
\end{equation}

where $\xi=\frac{r_{p}}{v\bar{t}^{\frac{1}{k}}}<1$ and
$\eta=\frac{t_{o}}{\bar{t}}>1$. The graphical  solution of
eqn.\ref{ovegrowth} is depicted in fig.6b and indicates that
$t_{o}$ (and therefore $\eta$) decreases with $k$. This is in
agreement with the results of fig.5 which shows that the
overgrowth phenomenon is more important at greater $k$.

As far as the guess function eqn.\ref{sim2} is concerned, it
matches the simulation curves with a very high degree of
correlation. For instance the output of the fit to the $n=2$
curve, gives $a=0.9750\pm 0.0007$, $b=0.088\pm 0.001$ and a
squared correlation coefficients practically $1$. For the sake of
completeness the $a$ and $b$ fitting parameters are shown in
fig.7, where $a$ is found to be nearly one. This is in agreement
with the theoretical value predicted by eqn.\ref{Uncov2}.

\begin{figure}[htbp]
\begin{center}
\includegraphics[scale=1]{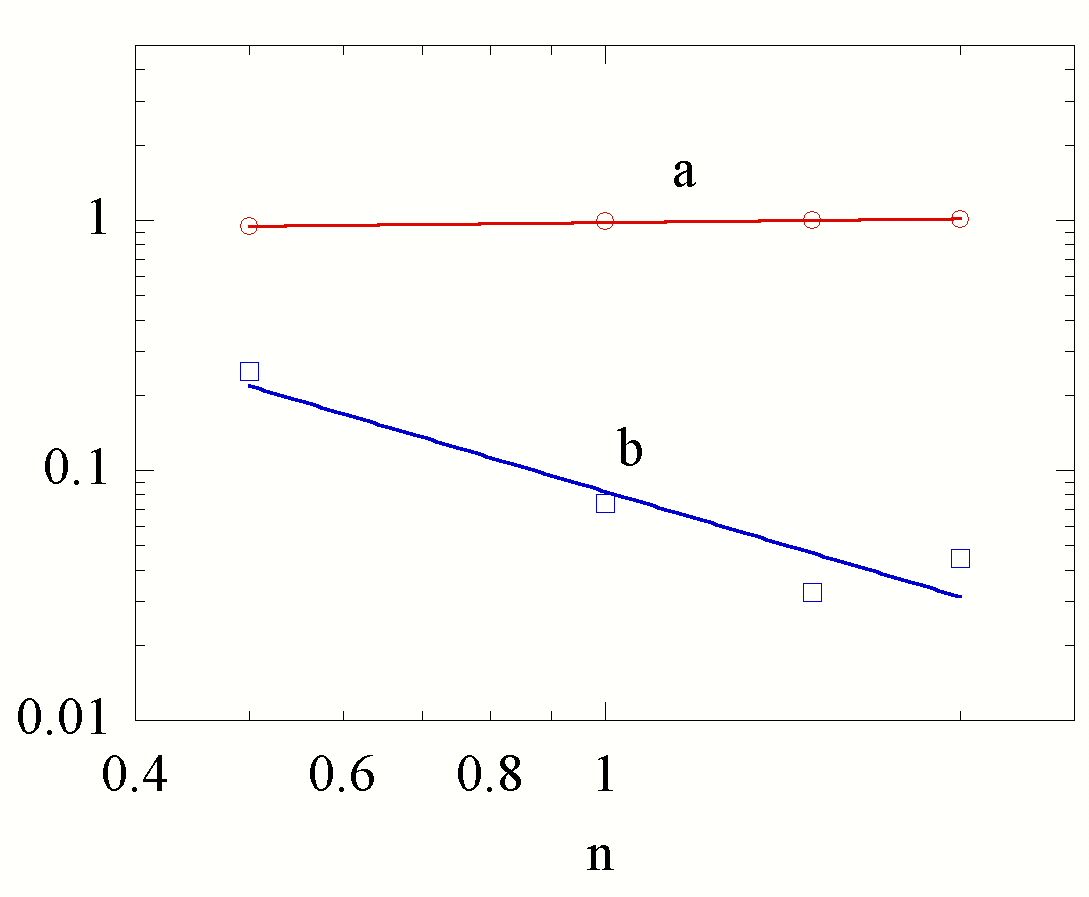}
\caption{Behavior of the fitting parameters a
and b of eqn.\ref{sim2} as a function of growth exponent $n$, for
KJMA-non-compliant growths.}
\end{center}
\end{figure}

The behavior of the transformed fraction for KJMA-compliant
growths are reported in fig.8 for $n=1$, $n=3/2$ and $n=2$. 

\begin{figure}[htbp]
\begin{center}
\includegraphics[scale=1]{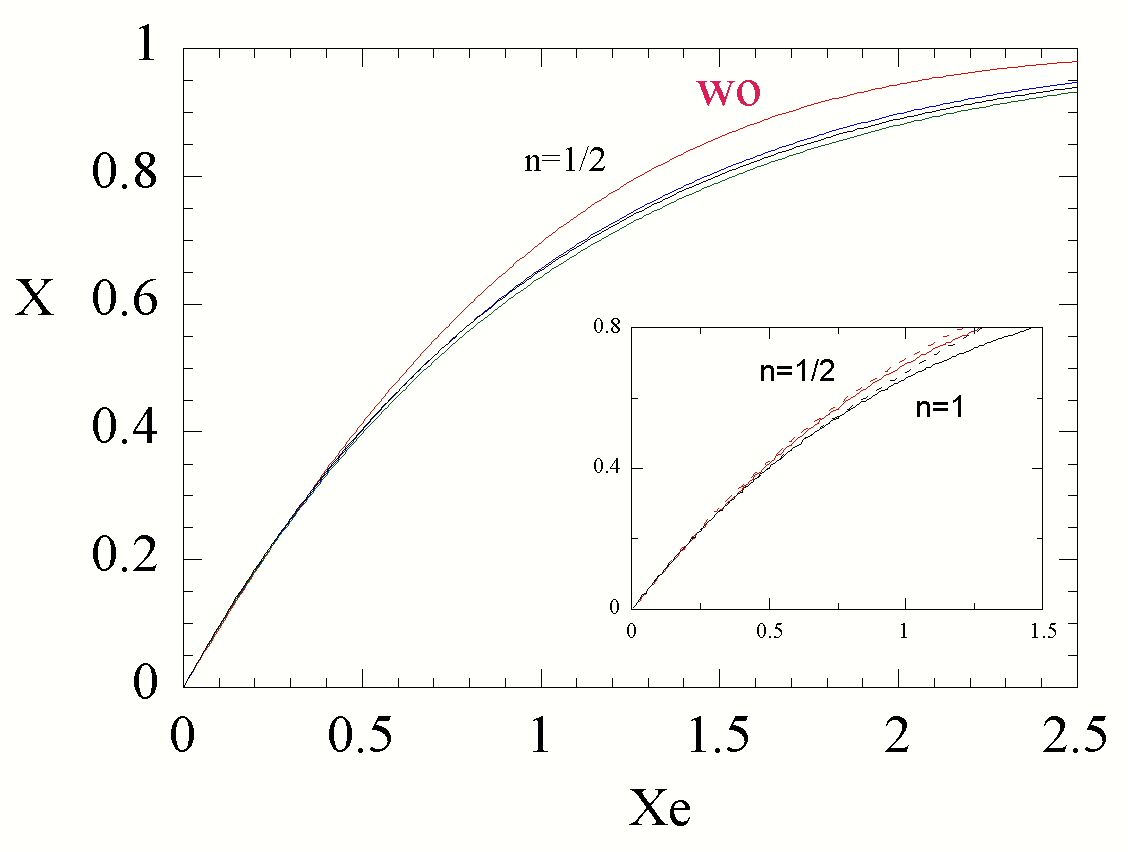}
\caption{ Kinetics of the actual surface
fraction, as a function of actual extended surface fraction, for
several values of $n$. In the graph the kinetics for $n=1/2$ is
compared with the KJMA-compliant growth with $n=1$, $n=2$ and
$n=3/2$ (from the top, respectively). In the inset the kinetics
for $n=1$ and $n=1/2$ are displayed together with the truncated
KJMA series expansions eqn.\ref{serie3} and eqn.\ref{parabola1},
respectively.}
\end{center}
\end{figure}
These
kinetics are very close to each other and differs, markedly, from
that at $n=1/2$ also reported in the same figure. In the inset,
the kinetics for $n=1/2$ and $n=1$ are compared with the KJMA
series expansion eqn.\ref{parabola1} and eqn.\ref{serie3},
respectively. The fact that the curves for $n=1$, $3/2$ and $2$
collapse on the same curve, can be rationalized computing the
coefficients of the series expansion of $Q(X_{e})$ for integer
$n$. In particular, by employing the method discussed in the
previous section the two last coefficients of the series (e.g.
eqn.\ref{serie3}) are $0.4960$, $0.1633$ and $0.5$, $0.1673$ for
$n=2$ and $3$, respectively. We also performed computer
simulations of phase transitions for non-constant actual
nucleation rate. The output of this computation is displayed in
fig.9 where the behavior of the nucleation rate is shown in the
inset as function of $X_{e}$. In particular, the actual nucleation
rate is given by the function $I_{a}(t)\approx
t^{2}\exp(-\tilde{a}t^{3})$.

\begin{figure}[htbp]
\begin{center}
\includegraphics[scale=1]{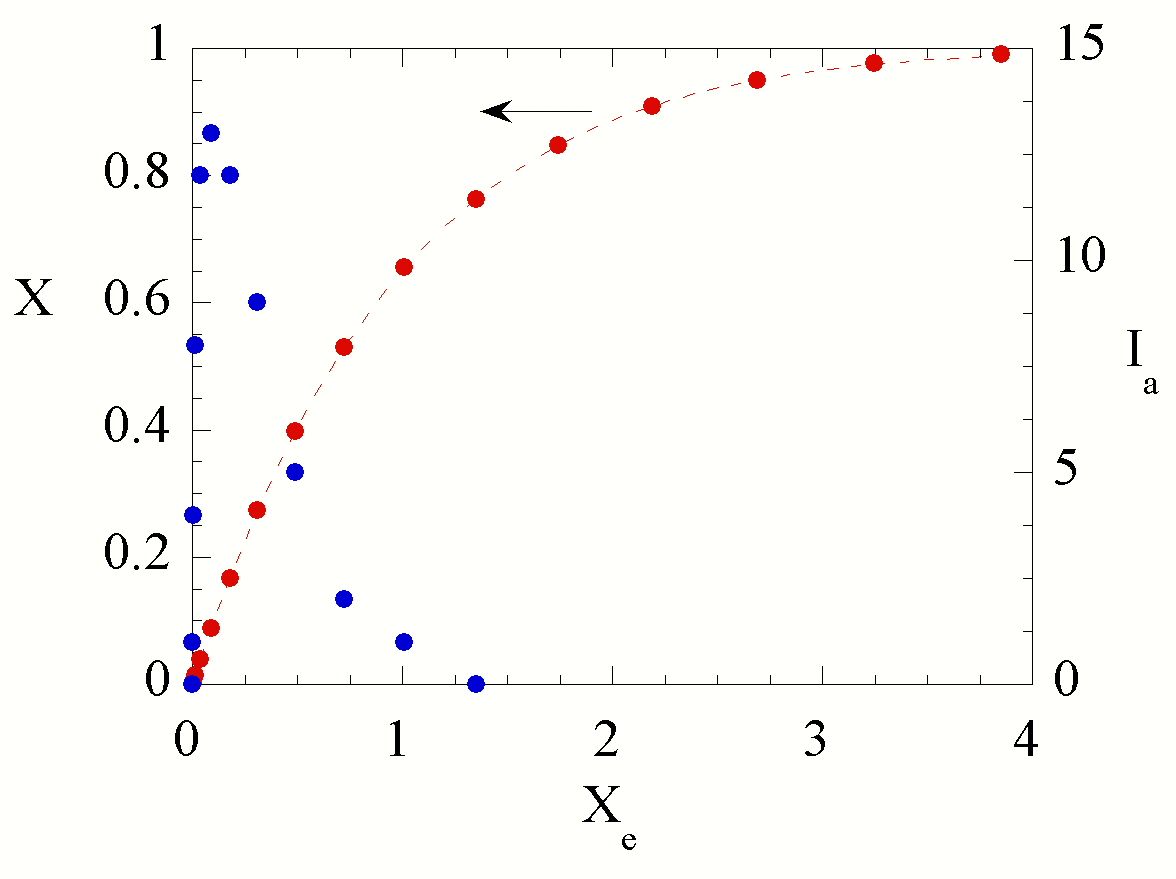}
\caption{Fig.9 Kinetics of the actual surface
fraction, as a function of the actual extended surface, for
non-constant actual nucleation rate (left scale). The best fit of
the function eqn.\ref{sim2} to the $X(X_{e})$  kinetics has been
shown as dashed line. The correlation coefficient of the fit is in
fact $1$ for the parameters $a=1.014$ and $ b=0.0382$. The actual
nucleation rate, as a function of $X_{e}$, is also reported on the
right scale in $nuclei \times 10^{6}/site$ units.}
\end{center}
\end{figure}

Also in this case the function
eqn.\ref{sim2} has been found to match the kinetics with high
degree of correlation where, again, the independent variable is
the actual extended surface fraction.

Let us address in more detail the question of the dependence of
volume fraction on extended volume fraction. To this end, we
discuss the second order term of the exact solution
eqn.\ref{Uncov1}, namely

\begin{equation}\label{changevar1}
\int_{0}^{t} I_{a}(t')dt'\int_{0}^{t'} I_{a}(t'')dt''
\int_{\Delta(t,t')}d\mathbf{r}_{1}A(r_{1},R(t',t''),R(t,t'')),
\end{equation}

where eqn.\ref{Area0} has been employed. We point out that in
eqn.\ref{changevar1} $A(r_{1},R(t',t''),R(t,t''))$ is a second
order homogeneous function of $r_{1}$, $R(t',t'')$ and $R(t,t'')$
variables. Accordingly,  for the growth law $R(t,t')=v(t-t')^n$,
using the re-scaled variables, $r'_{1}=r_{1}/vt^n$, $\tau'=t'/t$
and $\tau'' =t''/t$ the integral becomes

 \begin{equation}\label{changevar2}
\pi^{2}v^{4}t^{4n+2}\int_{0}^{1}
I_{a}(\tau')d\tau'\int_{0}^{\tau'} I_{a}(\tau'')d\tau''
\int_{\Delta(\tau')}d\mathbf{r}'_{1}A(r'_{1},\tau',\tau'').
\end{equation}

Eqn.\ref{changevar2} takes the form $C_{2}X_{e}^{2}$ where
\\$C_{2}(n,
[I_{a}(\tau)])=\frac{\int_{0}^{1}d\tau'I_{a}(\tau')\int_{0}^{\tau'}I_{a}(\tau'')
 d\tau''
\int_{\Delta(\tau')}A(r_{1}',\tau',\tau'')d\mathbf{r}'_{1}
}{(\int_{0}^{1}I_{a}(\tau')(1-\tau')^{2n}d\tau')^{2}}$ depends on
$n$ and actual nucleation rate. It is apparent that in the case
discussed in the previous section $I_{a}(t)=I_{a}$, $C_{2}(n)$, as
well as higher order coefficients, is a function of $n$, only. In
this case the transformed fraction is expected of the form
$1-X=\sum_{k} C_{k}(n)X_{e}^{k}$.

On the other hand, in the case of a constant "phantom-included"
nucleation rate, $I_{a}(t)=I_{0}(1-X(t))=I_{0}Q(t)$, and
eqn.\ref{Uncov1} becomes an integral equation for the $Q(t)$
unknown. With reference to the second order term, in this case
eqn.\ref{changevar2} takes the general form $C'(n)\hat{X}_{e}^{2}$
which now implies the series $1-X=\sum_{k}
C'_{k}(n)\hat{X}_{e}^{k}$ (note that  this is a series expansion
in terms of phantom-included extended surface). In the specific
case of KJMA-compliant growths, however, these series reduces to
the exponential series with constant coefficients
$\frac{(-1)^{k}}{k!}$. It is instructive to estimate the first two
coefficients in the case of linear growth. For constant phantom
included nucleation rate the untransformed fraction satisfies the
integral equation,

\begin{equation}\begin{array}{l}\label{Integraleq1}
Q(t)=1-I_{0}\int_{0}^{t}Q(t')|\Delta(t,t')|dt'+\\
\\
+I_{0}^{2}\int_{0}^{t}dt'\int_{0}^{t'}dt''Q(t')Q(t'')\int_{\Delta(t,t')}d\mathbf{r}_{1}A(r_{1},t,t',t'')+O(I_{0}^{3}).
\end{array}
\end{equation}

The first order term of this equation, namely of the order of
$I_{0}$, gives $Q(t)=1-I_{0}\pi v^{2}t^{3}/3=1-\hat{X}_{e}$. By
substituting $Q\approx 1-\hat{X}_{e}$ in eqn.\ref{Integraleq1}, we
get

\begin{equation}\begin{array}{l}\label{Integraleq2}
Q(t)=1-\hat{X}_{e}+I_{0}\int_{0}^{t}\hat{X}_{e}(t')|\Delta(t,t')|dt'+\\
\\ +I_{0}^{2}\int_{0}^{t}dt'\int_{0}^{t'}dt''\int_{\Delta(t,t')}d\mathbf{r}_{1}A(r_{1},t,t',t'')+O(I_{0}^{3}).
\end{array}
\end{equation}

Using dimensionless variables $r_{1}'=r_{1}/vt$, $\tau'=t'/t$ and
$\tau''=t''/t$ eqn.\ref{Integraleq2} eventually becomes

\begin{equation}\begin{array}{l}\label{Integraleq3}
Q(t)=1-\hat{X}_{e}+3\hat{X}_{e}^{2}\int_{0}^{1}\tau'^{3}(1-\tau')^{2}d\tau'+\\
\\+9\hat{X}_{e}^{2}\int_{0}^{1}d\tau'\int_{0}^{\tau'}d\tau''(1-\tau')^{2}A(\tau',\tau'')
+O(I_{0}^{3}),
\end{array}
\end{equation}

where $A(\tau',\tau'')$ is given through eqn.\ref{Aria1}. Notably,
the last term in eqn.\ref{Integraleq3} has been already estimated
in eqn.\ref{Secondterm} and is equal to
$\frac{9}{20}\hat{X}_{e}^{2}$. The coefficient of
$\hat{X}_{e}^{2}$ is eventually computed as
$\frac{3}{60}+\frac{9}{20}=\frac{1}{2}$, that is the expected
result.

It is worth noting that the present approach can also be applied
to different convex shapes other than circles and spheres,
provided the orientation of nuclei is the same (with a possible
exception for triangle) . This aspect has been discussed in
details in refs.\cite{bib:shape1},\cite{bib:shape2}.

We conclude this section by quoting the recent results of
ref.\cite{bib:Aleks}. In this noteworthy contribution the author
faced the problem of describing the kinetics in terms of the
actual nucleation rate. An ingenious application of the so called
"Differential critical region" approach makes it possible to find
the $Q(t)$ kinetics by solving an appropriate integral equation
\cite{bib:Aleks}. On the other hand, the different method employed
in the present work, based on the use of correlation function,
pertains to the same class of stochastic approaches on which
"Kolmogorov's method" is rooted. It could be enlightening, for the
present topic, to demonstrate that the two approaches are in fact
equivalent.

\section{Conclusions}
We have shown that, employing the correlation function approach,
the constraints on growth laws underlying the KJMA theory can be
eliminated. In other words, the present modeling is not
constrained to any form of the growth law. The actual extended
volume fraction is shown to be the natural variable of the
kinetics, which implies universal curves. Besides, we proposed a
formula to fit experimental data by using the measurable actual
extended coverage. The displacement of the kinetics from the
exponential law, i.e. the $b$ parameter in eqn.\ref{sim2}, may
give insights into the microscopic growth law of nuclei.

\end{document}